\documentclass[superscriptaddress,twocolumn,showpacs,preprintnumbers,amsmath,amssymb]{revtex4}
\usepackage{graphicx}
\usepackage{dcolumn}
\usepackage{bm}
\usepackage{latexsym}
\begin{document}
\title{Stochastic kinetics of a single headed motor protein:\\
dwell time distribution of KIF1A}

\author{Ashok Garai} 
\author{Debashish Chowdhury}
\affiliation{Department of Physics, Indian Institute of Technology,
           Kanpur 208016, India.  }

\begin{abstract}
KIF1A, a processive single headed kinesin superfamily motor, hydrolyzes 
Adenosine triphosphate (ATP) to move along a filamentous track called   
microtubule. The stochastic movement of KIF1A on the track is 
characterized by an alternating sequence of pause and translocation. 
The sum of the durations of pause and the following translocation 
defines the dwell time. Using the NOSC model (Nishinari et. al. PRL, 
{\bf 95}, 118101 (2005)) of individual KIF1A, we systematically derive 
an analytical expression for the dwell time distribution. More detailed 
information is contained in the probability densities of the ``conditional 
dwell times'' $\tau_{\pm\pm}$ in between two consecutive steps each of 
which could be forward (+) or backward (-). We calculate the probability 
densities $\Xi_{\pm\pm}$ of these four conditional dwell times. However, 
for the convenience of comparison with experimental data, we also present 
the two distributions $\Xi_{\pm}^{*}$ of the times of dwell before a 
forward (+) and a backward (-) step. In principle, our theoretical 
prediction can be tested by carrying out single-molecule experiments 
with adequate  spatio-temporal resolution. 
\end{abstract}
\pacs{87.16.ad, 87.16.Nn, 87.10.Mn}
\maketitle

\section{Introduction}

Molecular motors are nano-devices which perform mechanical work by 
converting part of the input energy; for the motors of our interest 
in this paper, the input is derived from the hydrolysis of ATP 
molecules \cite{b.howard01}. In reality, these motors are also enzymes 
that hydrolyze ATP and utilize the input chemical energy to perform 
mechanical work. In this paper we specifically consider the members 
of a particular superfamily of motors, called Kinesin, which are 
involved in intracellular transport processes in living cells. This 
family is designated as KIF1A and the members of this family move 
along filamentous tracks called microtubule (MT) \cite{hirokawa09}. 

One unique feature of a KIF1A is that, at least under the conditions 
of {\it in-vitro} experiments, it functions as a single-headed motor.  
The average properties, e.g., the average velocity, of these motors 
have been calculated analytically by using a theoretical model 
developed by Nishinari, Okada, Schadschneider and Chowdhury (from now 
onwards, referred to as the NOSC model) \cite{nosc,greulich07} which 
is an extension of the general approach pioneered by Fisher and 
Kolomeisky \cite{kolofish}. 

In single molecule experiments, individual motor proteins are observed 
to move in an alternating sequence of pause and translocation. The sum 
of the pause at a binding site and the subsequent translocation can 
be defined as the corresponding ``dwell time''. Because of the 
intrinsic irreversibility of mechano-chemical kinetics of the system, 
the inverse of the mean dwell time is the average velocity of a motor. 
The dwell time distribution $g(t)$ contains more detailed information 
on the stochastic kinetics of a motor than that revealed its average 
velocity. For example, the randomness parameter 
\begin{equation}
 r=\frac{<t^2>-<t>^2}{<t>^2}
\label{eq-r}
\end{equation}
provides an estimate of the lower bound on the number of rate-limiting 
kinetic steps in each cycle of the motor \cite{schnitzer95}.
For some other motors, which move on nucleic acid strand, the analytical 
forms of the distributions of the dwell times have been reported 
recently \cite{gccr,tsc09}.  

In this paper we report the exact analytical expression for the 
distribution of the dwell times of a KIF1A motor in the NOSC model 
during a single processive run in between its attachment to the track 
and the next detachment. What makes the calculation more difficult 
in the case of KIF1A, compared to those of those reported in 
ref.~\cite{gccr,tsc09}, is the occurrence of branched pathways in its 
mechano-chemical cycle. For motors which can step both forward and 
backward, one can define conditional dwell times which may be more 
easily extracted from the data obtained from single molecule experiments 
\cite{linden07,fisher07}. Therefore, in this paper we also report 
analytical expressions for the probability densities of these 
conditional dwell times as well as that of a few other closely related 
random variables.

\begin{figure}
 \begin{center}
  \includegraphics[width=0.85\columnwidth]{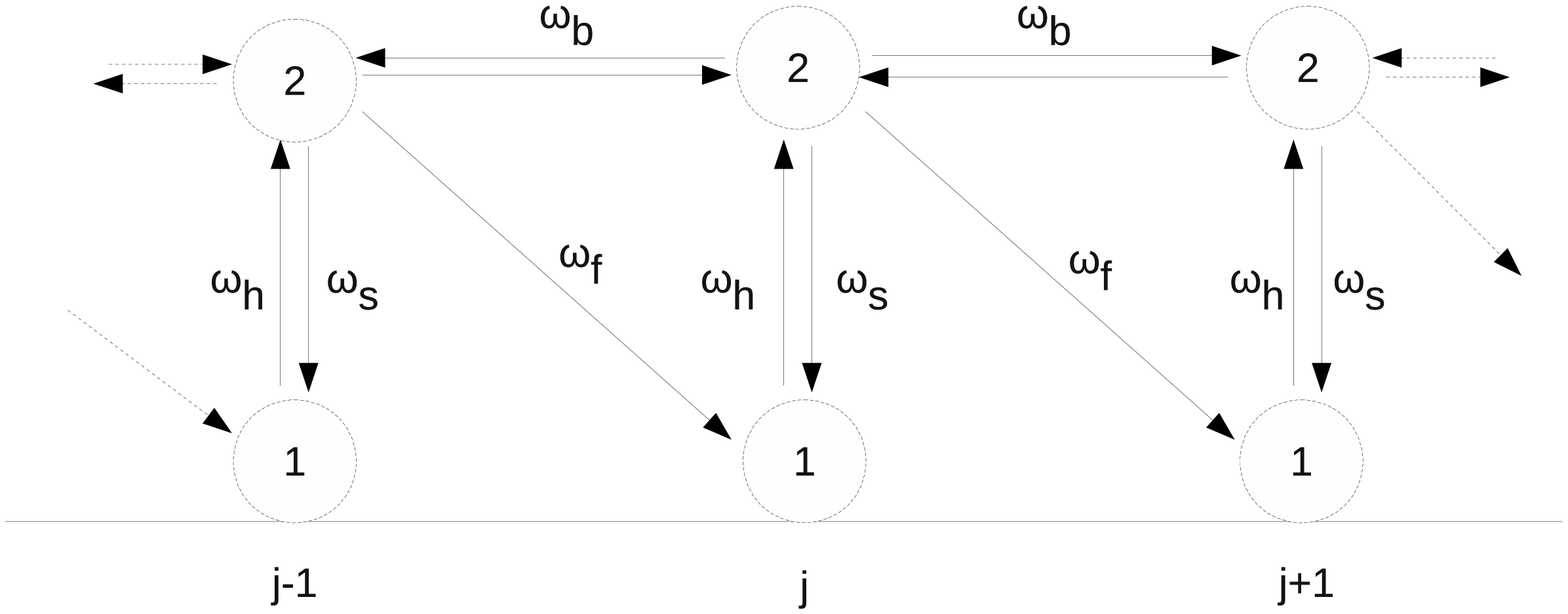}
 \end{center}
\caption{Two state model for KIF1A. The indices $...,j-1,j,j+1,...$ 
label the equispaced sites for the binding of the motor to its track. 
The states $1$ and $2$ correspond to the ``chemical'' states in which 
the motor is bound strongly and weakly, respectively, to the 
microtubule track. The allowed transitions are shown by the arrows 
along with the corresponding rate constants (transition probability 
per unit time).}
\label{fig-modelb}
\end{figure}

\section{Model and stochastic kinetics}

\subsection{The Model}

A MT track of the motor is modelled as a one-dimensional finite 
lattice having $L$ number of discrete sites. Each site corresponds 
to a KIF1A binding site on the MT and the lattice spacing is the 
separation between the successive binding sites on a MT. A KIF1A motor 
is represented by a particle with two possible chemical states labeled 
by the indices $1$ and $2$. The states $1$ and $2$ correspond to the 
strongly bound and weakly bound states, respectively. fig.~\ref{fig-modelb} 
illustrates the detailed mechano-chemical cycle of KIF1A. The 
transitions $1_{j} \leftrightarrow 2_{j}$ are purely chemical whereas 
the transitions $2_{j} \leftrightarrow 2_{j \pm 1}$, which correspond 
to the Brownian motion, are purely mechanical. In contrast, the transition 
$2_{j} \rightarrow 1_{j+1}$ is mechano-chemical (see ref. \cite{nosc} 
for a more detailed description).  

\subsection{Kinetics and the master equations}

We define $S(j,t)$ and $W(j,t)$ as the probabilities of finding KIF1A in state $1$ and state $2$ respectively, at site $j$ at time $t$. The master equations for these probabilities are 
\begin{equation}
 \frac{dS(j,t)}{dt}=-\omega_hS(j,t)+\omega_fW(j-1,t)+\omega_sW(j,t)
\label{eq-wbwfS}
\end{equation}
\begin{eqnarray}
\frac{dW(j,t)}{dt}&=&\omega_hS(j,t)-(\omega_f+\omega_s+2\omega_b)W(j,t) \nonumber \\
&+&\omega_b(W(j-1,t)+W(j+1,t))
\label{eq-wbwfW}
\end{eqnarray}
with $\omega_h=\omega^{0}_h [ATP]$, where $[ATP]$ is the concentration of ATP.
Now we introduce Fourier transforms of $S(j,t)$ and $W(j,t)$ by 
${\bar S}(q,t)=\sum_{j=-\infty}^{\infty}S(j,t)e^{-\imath qj}$ and ${\bar W}(q,t)=\sum_{j=-\infty}^{\infty}W(j,t)e^{-\imath qj}$, where $\imath=\sqrt{-1}$ and the lattice spacing, $d=1$. Thus, 
\begin{equation}
\left[ \begin{array}{c} \dot{{\bar S}}(q,t) \\ \dot{{\bar W}}(q,t) \end{array} \right] = 
M(q) \left[ \begin{array}{c} {\bar S}(q,t) \\ {\bar W}(q,t) \end{array} \right] \label{eq-matrix}
\end{equation}
where the transition matrix ${\bf M}(q)$ is given by,
\begin{equation}
{\bf M}(q)= \left( \begin{array}{cc}
-\omega_h & (\omega_fe^{-\imath q}+\omega_s) \\
\omega_h &  -\left\lbrace(2\omega_b+\omega_s+\omega_f)-\omega_b(e^{-\imath q}+e^{\imath q})\right\rbrace\end{array} \right).
\end{equation}
The Laplace transforms of the Fourier transforms of ${\bar S}(q,t)$ and 
${\bar W}(q,t)$  are given by ${\tilde S}(q,s)=\int_{0}^{\infty}dte^{-st}{\bar S}(q,t)$ 
and ${\tilde W}(q,s)=\int_{0}^{\infty}dte^{-st}{\bar W}(q,t)$. 
Therefore, using ${\tilde S}(q,s)$ and ${\tilde W}(q,s)$ as the two components of a  
vector ${\bf {\tilde P}}(q,s)$, the matrix equation (\ref{eq-matrix}) can now 
be written in a more compact notation as  
\begin{equation}
 {\bf {\tilde P}}(q,s)={{\bf R}(q,s)}^{-1} {\bf P}(0), {\rm with} 
 {\bf R}(q,s)\equiv s{\bf I}-{\bf M}(q)
\label{eq-prob-den}
\end{equation}
and ${\bf P}(0)$ is a vector whose elements are determined by the 
initial conditions for $S$ and $W$.  The initial conditions are 
$S(j,0)=\delta_{jk}$ and $W(j,0)=0$ where $k$ is an arbitrarily 
selected site. The dwell time at the site $k$ is the total duration  
for which the motor stays at that site, starting from the initial 
condition mentioned above, irrespective of the chemical state, 
before its next departure from the same site.

Thus the determinant of ${\bf R}(q,s)$ is a $2$nd order polynomial in 
$s$; that is \cite{chemla08},
\begin{equation}
 |{\bf R}|(q,s)=s^2+\alpha(q)s+\gamma(q)
\end{equation}
where,
\begin{equation}
 \alpha(q)=\omega_h+2\omega_b+\omega_s+\omega_f-\omega_b(e^{-\imath q}+e^{\imath q})
\end{equation}
\begin{equation}
 \gamma(q)=\omega_h(2\omega_b+\omega_f)-\omega_b\omega_h(e^{-\imath q}+e^{\imath q})-\omega_h\omega_fe^{-\imath q}
\end{equation}
Note that $\gamma(q)$ is the determinant of the transition matrix $M(q)$. Because of conservation of probability, in the $q\rightarrow0$ limit, all of the columns of ${\bf M}$ sum to zero. Therefore, we obtain $|{\bf M}(0)|=\gamma(0)=0$. 
We define \cite{chemla08} the position probability density ${\bar P}(q,t) = {\bar S}(q,t) + {\bar W}(q,t)$, and hence 
\begin{equation}
 {\tilde P}(q,s)=\frac{s+\alpha(0)}{s^2+\alpha(q)s+\gamma(q)}.
\end{equation}

\subsection{Velocity and Diffusion constant}
Following ref.\cite{chemla08}, the average velocity $v$ of KIF1A and the diffusion constant $D$ are found to be 
\begin{equation}
 v=-\imath \frac{\dot{\gamma}(0)}{\alpha(0)}=\frac{\omega_f\omega_h}{(\omega_f+\omega_h+\omega_s)}.
\label{eq-kifsgv}
\end{equation}
and 
\begin{equation}
 D=\frac{\ddot {\gamma}(0)-2\imath v\dot{\alpha}(0)-2\beta(0)v^2}{2\alpha(0)}=\frac{2\omega_b\omega_h+\omega_f\omega_h-2v^2}{2(\omega_f+\omega_h+\omega_s)},
\label{eq-kifsgD}
\end{equation}
where $\dot{\gamma}(0)$ and $\ddot{\gamma}(0)$ are the first and second derivatives, respectively, of $\gamma(q)$ with respect to $q$ evaluated at $q=0$ while 
$\beta(0)$ is the coefficient of $s^2$ at $q=0$; in this case $\beta(0)=1$. Interestingly, the velocity $v$ 
depends only on the coefficients $\gamma$ and $\alpha$, of the lowest 
two orders of the polynomial obtained from the determinant of 
${\bf R}(q,s)$. On the other hand, the diffusion coefficient $D$ depends 
on the three lowest order coefficients $\beta$, $\alpha$, and $\gamma$ 
of the determinant of ${\bf R}(q,s)$.

\begin{figure}[t]
 \begin{center}
  \includegraphics[width=0.85\columnwidth]{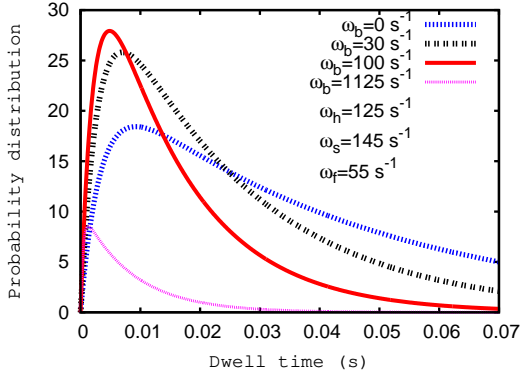}
 \end{center}
\caption{Dwell time distribution is plotted from equation (\ref{eq-dwellbranchkif}) for few different values of $\omega_b$.}
\label{fig-dwellwbwfkifs}
\end{figure}

\section{Dwell time distribution}
We define the individual forward and backward branching probabilities 
as $p_{+}, p^{'}$ and $p_{-}$ and the corresponding dwell time 
distributions are $\psi_{+}, \psi^{'}~~\text{and}~~\psi_{-}$, 
respectively. Thus 
\begin{equation}
 p_{+}=\frac{\omega_b}{2\omega_b+\omega_f},
\end{equation}
\begin{equation}
 p^{'}=\frac{\omega_f}{2\omega_b+\omega_f},
\end{equation}
and,
\begin{equation}
 p_{-}=\frac{\omega_b}{2\omega_b+\omega_f}.
\end{equation}
We define the Fourier weights $\rho_{+}(q)=e^{-\imath q}$ and 
$\rho_{-}(q)=e^{\imath q}$ for the forward and backward steps, 
respectively (step size $d=1$ in our units). Hence \cite{chemla08},
\begin{equation}
 \frac{1}{s{\tilde P}(q,s)}|_{\left\lbrace \rho_{k\neq k^{'}}(q)=0 \right\rbrace} =\frac{1-\zeta(s)\rho_{+}(q)-p_{-}\psi_{-}(s)\rho_{-}(q)}{1-\zeta(s)-p_{-}\psi_{-}(s)},
\label{s.long}
\end{equation}
where $\zeta(s)=(p_{+}\psi_{+}(s)+p^{'}\psi^{'}(s))$ and the symbol 
${\left\lbrace \rho_{k\neq k^{'}}(q)=0 \right\rbrace}$ expresses the condition 
that the Fourier weight for all possible steps, except $k^{'}$ is zero;  
in our case, $k^{'}$ can represent either the forward ($+$) or the backward 
($-$) steps. 
For forward branching we obtain,
\begin{eqnarray}
 \frac{1}{s{\tilde P}(q,s)}\mid_{\left\lbrace \rho_{-}(q)=0 \right\rbrace} &=&\frac{1-p_{+}\psi_{+}(s)\rho_{+}(q)-p^{'}\psi^{'}(s)\rho_{+}(q)}{1-p_{+}\psi_{+}(s)-p^{'}\psi^{'}(s)}\nonumber \\
&=&a_{0}+a_{+}(s)\rho_{+}(q)
\end{eqnarray}
where 
\begin{eqnarray}
 a_0&=&\frac{s^2+(2\omega_b+\omega_h+\omega_f+\omega_s)s+\omega_h(2\omega_b+\omega_f)}{s^2+(\omega_h+\omega_f+\omega_s)s} \nonumber \\
 a_{+}&=&-\frac{\omega_bs+\omega_h(\omega_b+\omega_f)}{s^2+(\omega_h+\omega_f+\omega_s)s}
\label{eq-a0}
\end{eqnarray}

Thus,
\begin{eqnarray}
&& p_{+}\psi_{+}(s)+p^{'}\psi^{'}(s)=-\frac{a_{+}(s)}{a_{0}}\nonumber \\
&=&\frac{\omega_bs+\omega_h(\omega_b+\omega_f)}{s^2+(2\omega_b+\omega_h+\omega_f+\omega_s)s+{\omega_h(2\omega_b+\omega_f)}}
\end{eqnarray}
Inverse Laplace transform yields,
\begin{equation}
 p_{+}\psi_{+}(t)+p^{'}\psi^{'}(t)=\frac{c_{1}e^{-r_1t/2}-c_{2}e^{-r_2t/2}}{2r_{0}}
\end{equation}
where,
\begin{equation}
 r_{0}=\sqrt{(2\omega_b+\omega_f+\omega_h+\omega_s)^2-4(2\omega_b+\omega_f)\omega_h}
\label{eq-r0}
\end{equation}
\begin{equation}
 c_1=(2\omega_b^2+\omega_b\omega_f-\omega_b\omega_h-2\omega_f\omega_h+\omega_b\omega_s+\omega_br_{0})
\end{equation}
\begin{equation}
 c_{2}=(2\omega_b^2+\omega_b\omega_f-\omega_b\omega_h-2\omega_f\omega_h+\omega_b\omega_s-\omega_br_{0})
\end{equation}
\begin{equation}
 r_1=2\omega_b+\omega_f+\omega_h+\omega_s+r_{0}
\label{eq-r1}
\end{equation}
\begin{equation}
 r_2=2\omega_b+\omega_f+\omega_h+\omega_s-r_{0}
\label{eq-r2}
\end{equation}
Similarly,
\begin{eqnarray}
 \frac{1}{s{\tilde P}(q,s)}|_{\left\lbrace \rho_{+}(q)=0\right\rbrace }&=&a_{0}+a_{-}(s)\rho_{-}(q)
\end{eqnarray}
where $a_0$ is given by eq.~(\ref{eq-a0}) and  
\begin{equation}
 a_{-}(s)=-\frac{\omega_bs+\omega_b\omega_h}{s^2+(\omega_f+\omega_h+\omega_s)s}
\end{equation}

\begin{eqnarray}
&& p_{-}\psi_{-}(s)=-\frac{a_{-}(s)}{a_{0}} \nonumber \\
&=&\frac{\omega_bs+\omega_b\omega_h}{s^2+(\omega_h+2\omega_b+\omega_s+\omega_f)s+\omega_h(2\omega_b+\omega_f)}.
\end{eqnarray}
Inverse Laplace transform yields,
\begin{equation}
 p_{-}\psi_{-}(t)=\frac{\omega_b\left\lbrace (r_1-2\omega_h)e^{-r_1t/2}-(r_2-2\omega_h)e^{-r_2t/2}\right\rbrace}{2r_{0}}
\end{equation}
Now total dwell time distribution can be written as follows:
\begin{eqnarray}
 g(t)&=&p_{+}\psi_{+}(t)+p^{'}\psi^{'}(t)+p_{-}\psi_{-}(t)\nonumber \\
&=&\frac{\lbrace \omega_b(r_2-2\omega_h)-\omega_f\omega_h \rbrace(e^{-r_1t/2}-e^{-r_2t/2})}{r_{0}}
\label{eq-dwellbranchkif}
\end{eqnarray}
The total dwell distribution is plotted in the fig.~\ref{fig-dwellwbwfkifs}. 
We observe that with the increase of $\omega_b$ the most probable of 
dwell time shifts towards a smaller value. This is because rate constant 
$\omega_b$ is related to diffusion of KIF1A in state $2$. For smaller 
values of $\omega_b$  dwell time is dominated by the rate $\omega_f$, 
but with the increase of rate $\omega_b$ the dwell time depends on 
both $\omega_b$ and $\omega_f$ and eventually the mean dwell time decreases. 

\section{Probabilities of splitting and conditional dwell times}

A typical trajectory of a KIF1A motor consists of a random sequence of 
forward and backward steps. Therefore, we can define four different 
{\it conditional dwell times} $\tau_{\pm\pm}$. Here $\tau_{++}$ is the 
dwell time in between two consecutive forward steps whereas $\tau_{--}$ 
is the dwell time in between two consecutive backward steps. Similarly, 
$\tau_{+-}$ is the dwell time in between two consecutive steps of which  
the first is forward and the second is backward whereas the opposite if 
true in case of $\tau_{-+}$. 

We denote the probability density functions for the conditional dwell 
times by the symbols $\Xi_{\pm\pm}(t)$. The integrated probabilities 
obtained from these probability densities are given by 
\begin{equation}
\mathcal{P}_{\pm\pm}(t)=\int _{0}^{t}\Xi_{\pm\pm}(t') dt'
\end{equation}
Obviously, 
$\lim_{t\rightarrow \infty}\mathcal{P}_{\pm\pm}(t)=\int _{0}^{t}\Xi_{\pm\pm}(t') dt'=1$.
Moreover, we introduce the ``pairwise splitting'' probabilities 
$\Pi_{\pm\pm}$ where $\Pi_{++}$ and $\Pi_{+-}$ represent the 
probability that a forward step is followed by a forward step or a 
backward step, respectively. Similarly, $\Pi_{-+}$ and $\Pi_{--}$ 
denote the probability that a backward step is followed by a forward 
step or a backward step, respectively. For the analysis of the 
experimental data and comparison with theoretical predictions, it 
is sometimes more convenient to divide the dwell times into two 
groups depending on the {\it direction of the following step}. 
We use the symbols $\Xi_{+}^*(t)$ and $\Xi_{-}^*(t)$ to denote the 
probability density functions for the dwell times before a forward 
$(+)$ and a backward $(-)$ steps, respectively. Obviously, 
$\Xi_{\pm}^*(t)$ are given by 
\cite{linden07}
\begin{eqnarray}
\Xi_{+}^*(t)&=&\Pi_{++}\Xi_{++}(t)+\Pi_{-+}\Xi_{+-}(t) \nonumber \\
\Xi_{-}^*(t)&=&\Pi_{+-}\Xi_{-+}(t)+\Pi_{--}\Xi_{--}(t)
\label{eq-Xistar}
\end{eqnarray}
In this section we derive analytical expressions for $\Pi_{\pm\pm}$,  
$\Xi_{\pm\pm}(t)$, and hence, $\Xi_{\pm}^*(t)$ following the 
procedure adopted in ref.\cite{linden07,fisher07}.

In our model, immediately after a forward step the KIF1A motor can 
be found in either of the two states ($Q_{\mu}(t); Q_1(t)=S(t)$ i.e. 
strongly bound state, and $Q_2(t)=W(t)$ i.e. weakly bound state) 
whereas it can exist only in state $2$ immediately after a backward 
step. The first escape problem for our model is governed by a 
reduced Master equation (as described in refs. \cite{linden07} and 
\cite{van92}), with the transition matrix
\begin{equation}
\Upsilon = \left( \begin{array}{cc}
-\omega_h & \omega_s \\
\omega_h & -(\omega_f+\omega_s+2\omega_b) \end{array} \right).
\end{equation}

The eigen values of $\Upsilon$ are 
\begin{equation}
 \lambda_{1,2}=\frac{-u_1\pm\sqrt{(\omega_h-\omega_f-\omega_s-2\omega_b)^2+4\omega_h\omega_s}}{2}
\label{eq-lambda12}
\end{equation}
where, $u_1=\omega_h+\omega_f+\omega_s+2\omega_b$ and define $u_2=\lambda_1\lambda_2=\omega_f\omega_h+2\omega_b\omega_h$.
The initial conditions, describing the distribution of states just after a $\pm$ (forward and backward) step, are given by
\begin{equation}
 Q^{+}_{\mu}(0)=\frac{\omega_f\delta_{\mu 1}+\omega_b\delta_{\mu 2}}{\omega_b+\omega_f} ~ \textrm{and}~ Q^{-}_{\mu}(0)=\delta_{\mu 2} 
\label{eq-ini}
\end{equation}
where $\mu=1,2$, $Q_1=S$ and $Q_2=W$ as described above.

Using the expressions for the probability currents associated with 
the allowed transitions in our model, we get  
\begin{eqnarray}
 \Pi_{\pm+}\mathcal{P}_{\pm+}(t) &=& \int^t_0 dt' \{\omega_b Q_2(t') + \omega_f Q_2(t')\} \nonumber \\
 \Pi_{\pm-}\mathcal{P}_{\pm-}(t) &=& \int^t_0 dt' \{\omega_b Q_2(t') \} 
\label{eq-dt2}
\end{eqnarray}

Following Lind\'{e}n and Wallin \cite{linden07}, by using the Ansatz 
\begin{equation}
 \mathcal{P}_{++}(t)=\mathcal{P}_{--}(t)=1+\sum_{\mu=1}^2\varepsilon_{\mu}e^{\lambda_{\mu}t} 
\end{equation}
 together with eqs.~(\ref{eq-ini})-(\ref{eq-dt2}) to compute $\partial_t\Pi_{\pm+}\mathcal{P}_{\pm+}(t)$ and 
$\partial_t\Pi_{\pm-}\mathcal{P}_{\pm-}(t)$ we obtain the following systems of linear equations: 
\begin{eqnarray}
\left( \begin{array}{cc}
1 & 1 \\
\lambda_1 & \lambda_2 \end{array} \right)\left( \begin{array}{cccc}
 \varepsilon^{++}_1& \varepsilon^{+-}_1 & \varepsilon^{-+}_1 & \varepsilon^{--}_1 \\
\varepsilon^{++}_2& \varepsilon^{+-}_2 & \varepsilon^{-+}_2 & \varepsilon^{--}_2 \end{array} \right) \nonumber \\
= \left( \begin{array}{cccc}
-1 & -1 & -1 & -1 \\
\frac{\omega_b}{\Pi_{++}} & \frac{\omega_b^2}{\Pi_{+-}(\omega_b+\omega_f)} & \frac{\omega_b+\omega_f}{\Pi_{-+}} & \frac{\omega_b}{\Pi_{--}} \end{array} \right).
\end{eqnarray}

Solving this and using $\Xi_{\pm\pm}(t)=\partial_t\mathcal{P}_{\pm\pm}(t)$ 
we obtain the following distributions of the conditional dwell times:
\begin{eqnarray}
 \Xi_{++}(t)&=& u_2 \frac{(e^{\lambda_1t}-e^{\lambda_2t})}{\lambda_1-\lambda_2}
+\frac{\omega_b}{\Pi_{++}}
\frac{(\lambda_1e^{\lambda_1t}-\lambda_2e^{\lambda_2t})}{(\lambda_1-\lambda_2)} \nonumber \\
 \Xi_{+-}(t)&=& u_2 \frac{(e^{\lambda_1t}-e^{\lambda_2t})}{\lambda_1-\lambda_2}
+\frac{\omega_b^2 (\lambda_1e^{\lambda_1t}-\lambda_2e^{\lambda_2t})}{\Pi_{+-} (\omega_b+\omega_f)(\lambda_1-\lambda_2)}
\nonumber \\
 \Xi_{-+}(t)&=& u_2 \frac{(e^{\lambda_1t}-e^{\lambda_2t})}{\lambda_1-\lambda_2}
+\frac{(\omega_b+\omega_f)}{\Pi_{-+}} \frac{(\lambda_1e^{\lambda_1t}-\lambda_2e^{\lambda_2t})}{(\lambda_1-\lambda_2)} \nonumber \\
 \Xi_{--}(t)&=& u_2 \frac{(e^{\lambda_1t}-e^{\lambda_2t})}{\lambda_1-\lambda_2}
+\frac{\omega_b}{\Pi_{--}} \frac{(\lambda_1e^{\lambda_1t}-\lambda_2e^{\lambda_2t})}{(\lambda_1-\lambda_2)}
\label{eq-Xi}
\end{eqnarray}
Hence  
\begin{equation}
 \Upsilon^T\left( \begin{array}{cc}
\Pi_{1-} & \Pi_{1+} \\
\Pi_{2-} & \Pi_{2+} \end{array} \right)
= -\left( \begin{array}{cc}
0 & 0 \\
\omega_b & (\omega_f+\omega_b) \end{array} \right).
\end{equation}

To derive the pairwise splitting probabilities we first solve for the $\Pi_{\mu\pm}$ and then weight them 
according to the initial conditions as of eq.~(\ref{eq-ini}):
\begin{equation}
 \Pi_{\pm+}=\sum_{\mu}\Pi_{\mu+}Q^{(\pm)}_{\mu}(0),  \Pi_{\pm-}=\sum_{\mu}\Pi_{\mu-}Q^{(\pm)}_{\mu}(0)
\end{equation}

We finally get the following splitting probabilities
\begin{eqnarray}
\Pi_{++}=\Pi_{-+}&=&\frac{\omega_h(\omega_f+\omega_b)}{\omega_h(\omega_f+2\omega_b)}\nonumber \\
\Pi_{--}=\Pi_{+-}&=&\frac{\omega_h\omega_b}{u_2}
\label{eq-Pi}
\end{eqnarray}
Thus, probability of a forward step is 
$[\omega_h(\omega_f+\omega_b)]/[\omega_h(\omega_f+2\omega_b)]$ 
irrespective of the direction of the preceeding step. Similarly, the 
probability of a backward step is 
$(\omega_h \omega_b)/[\omega_h(\omega_f+2\omega_b)]$  
irrespective of the direction of the preceeding step. Both these are 
consistent with the kinetic pathways shown in fig.~\ref{fig-modelb} 
as well as with the fact that $\Pi_{++}+\Pi_{+-}=1$ and $\Pi_{--}+\Pi_{-+}=1$.

\begin{figure}[t]
 \begin{center}
  \includegraphics[width=0.85\columnwidth]{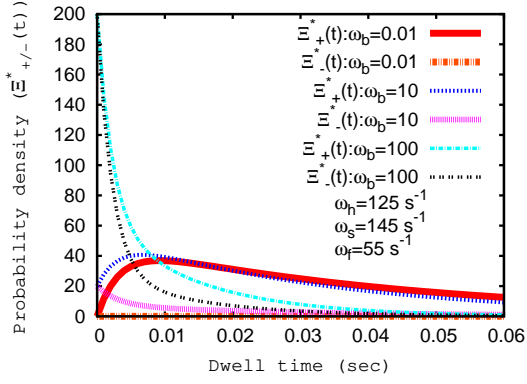}
 \end{center}
\caption{The distribution $\Xi_{\pm}^*(t)$ is plotted for few different 
values of $\omega_b$.}
\label{fig-Xistar}
\end{figure}

Substituting the expressions (\ref{eq-Pi}) and (\ref{eq-Xi}) into  
(\ref{eq-Xistar}) we get the analytical expressions for $\Xi_{\pm}^*(t)$.
This distribution is plotted in fig.~\ref{fig-Xistar} for a few 
different values of $\omega_b$. The most probable dwell time before 
a forward step decreases with increasing $\omega_b$.

\section{Randomness parameter}
\label{sec-randpar}

\begin{figure}
 \begin{center}
  \includegraphics[width=0.85\columnwidth]{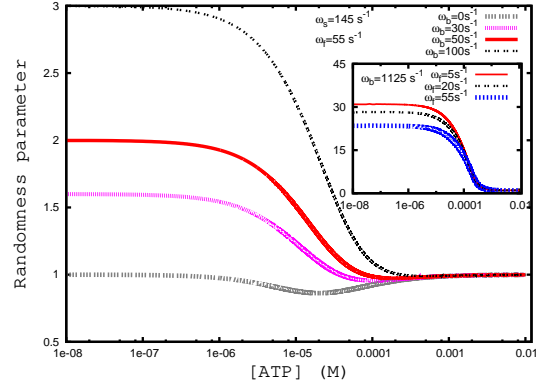}
 \end{center}
\caption{The randomness parameter $r$, defined by eq.~(\ref{eq-for28}), is plotted against ATP concentration for a few different values of $\omega_b$. The inset shows the same for few different values of $\omega_f$.}
\label{fig-1new}
\end{figure}

Using eq.~(\ref{eq-dwellbranchkif}) we obtain
\begin{equation}
 <t>=4\alpha\frac{(r_2+r_1)(r_2-r_1)}{(r_1r_2)^2}
\label{eq-28a}
\end{equation}
and
\begin{equation}
 <t^2>=16\alpha\frac{(r_2-r_1)(r_2^2+r_1r_2+r_1^2)}{(r_1r_2)^3}
\label{eq-28b}
\end{equation}
where $r_1$, $r_2$ are given by eqs.~(\ref{eq-r1}), (\ref{eq-r2}) and 
$\alpha$ is given by
\begin{equation}
 \alpha=\frac{2\omega_b^2+\omega_b\omega_f+\omega_b\omega_s-\omega_b\omega_h-\omega_f\omega_h-\omega_br_0}{r_0}.
\end{equation}
$r_0$ is given by eq.~(\ref{eq-r0}).
Using equations (\ref{eq-r}), (\ref{eq-28a}) and (\ref{eq-28b}) we obtain randomness parameter as follows
\begin{equation}
 r=\frac{r_1r_2(r_2^2+r_1r_2+r_1^2)-\alpha(r_2-r_1)(r_2+r_1)^2}{\alpha(r_2-r_1)(r_1+r_2)^2}
\label{eq-for28}
\end{equation}
Randomness parameter (see eq.~(\ref{eq-for28})) is plotted in fig.~\ref{fig-1new} against ATP concentration for
a few different values of $\omega_b$ and same is plotted in the inset for a few different values of $\omega_f$. 
With increasing concentration of ATP, $r$ decreases and finally sturates near to unity. At low ATP concentrations 
$r$ is greater than $1$ which may be the effect of multi-exponentiality in dwell time distribution. It also depicts 
that increasing concentration of ATP reduces the fluctuations in dwell time.  

\begin{figure}
 \begin{center}
  \includegraphics[width=0.85\columnwidth]{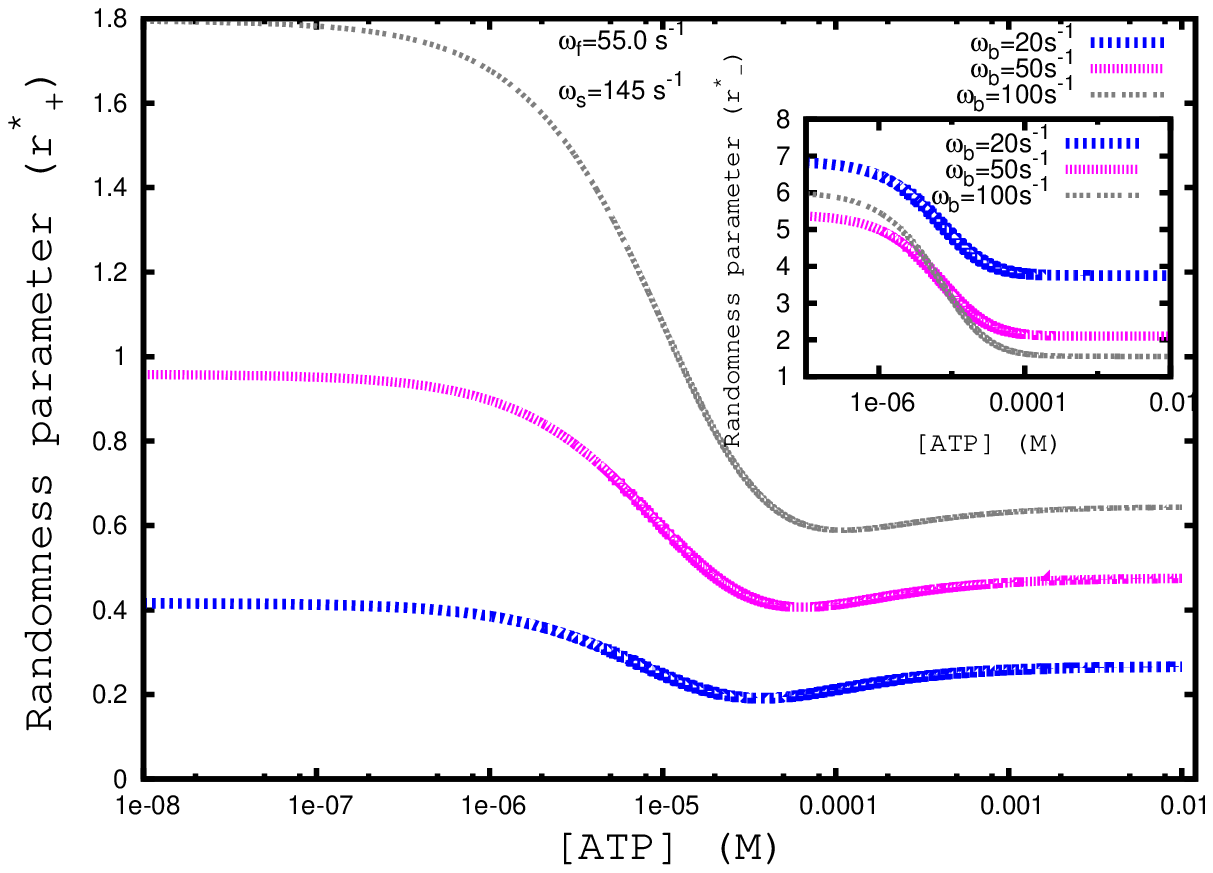}
 \end{center}
\caption{ The randomness parameter $r^{*}_+$, defined by eq.~(\ref{eq-for30a}), is plotted against ATP concentration for a few different values of $\omega_b$. The inset depicts the dependence of randomness parameter $r^{*}_{-}$, defined by eq.~(\ref{eq-for30a}), against ATP concentration for the same set of $\omega_b$ values.}
\label{fig-2new}
\end{figure}
\begin{figure}
 \begin{center}
 \includegraphics[width=0.85\columnwidth]{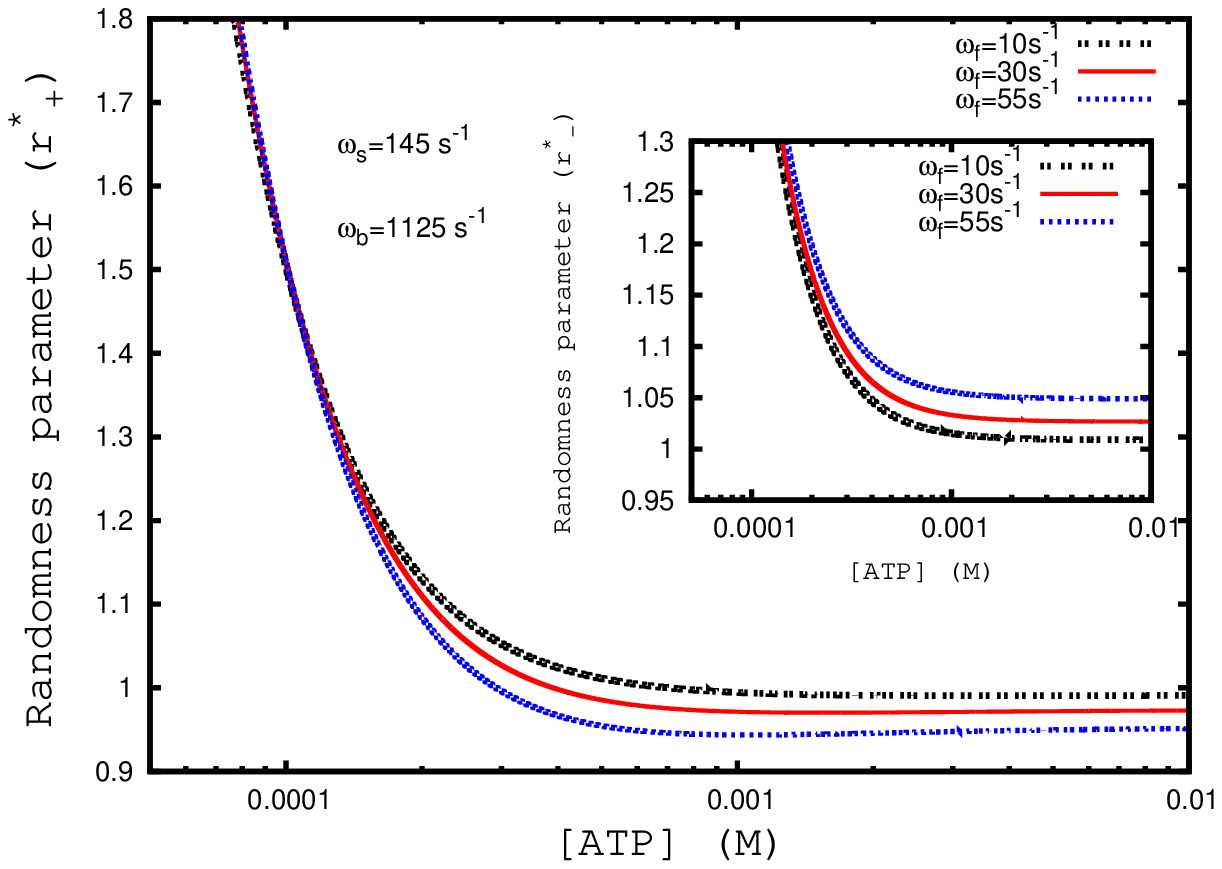}
 \end{center}
\caption{ The randomness parameter $r^{*}_+$, defined by eq.~(\ref{eq-for30a}), is plotted against ATP concentration for a few different values of $\omega_f$. The inset depicts the dependence of randomness parameter $r^{*}_{-}$, defined by eq.~(\ref{eq-for30a}), against ATP concentration for the same set of $\omega_f$ values.}
\label{fig-3new}
\end{figure}

For the conditional dwell times, the randomness parameters are defined by  
\begin{equation}
 r^*_{\pm}=\frac{<(t^*_{\pm})^2>-<t^*_{\pm}>^2}{<t^*_{\pm}>^2}
\label{eq-for30a}
\end{equation}
Using eq.~(\ref{eq-Xistar}) we obtain 
\begin{equation}
 <t^*_+>=-2\Pi_{++}\frac{u_2(\lambda_1+\lambda_2)}{(\lambda_1\lambda_2)^2}-\frac{\Pi_{++}\omega_b^2}{\Pi_{+-}(\omega_b+\omega_f)\lambda_1\lambda_2}-\frac{\omega_b}{\lambda_1\lambda_2}
\end{equation}
and
\begin{eqnarray}
<(t^*_+)^2> &=& 4\Pi_{++}\frac{u_2(\lambda^2_1+\lambda_1\lambda_2+\lambda^2_2)}{(\lambda_1\lambda_2)^3}+\frac{2\omega_b(\lambda_1+\lambda_2)}{(\lambda_1\lambda_2)^2} \nonumber \\
&&+2\Pi_{++}\frac{\omega_b^2(\lambda_1+\lambda_2)}{\Pi_{+-}(\omega_b+\omega_f)(\lambda_1\lambda_2)^2}
\end{eqnarray}
where $\Pi_{++}, \Pi_{-+}, \Pi_{--},$ and $\Pi_{+-}$ are given by 
eq.~(\ref{eq-Pi}). 
$\lambda_1$ and $\lambda_2$ are given by equation $(\ref{eq-lambda12})$. 
Similarly, 
\begin{equation}
<t^*_->=-2\Pi_{--}\frac{u_2(\lambda_1+\lambda_2)}{(\lambda_1\lambda_2)^2}-\frac{\Pi_{+-}(\omega_b+\omega_f)}{\Pi_{-+}\lambda_1\lambda_2}-\frac{\omega_b}{\lambda_1\lambda_2}
\end{equation}
and
\begin{eqnarray}
<(t^*_+)^2> &=& 4\Pi_{--}\frac{u_2(\lambda^2_1+\lambda_1\lambda_2+\lambda^2_2)}{(\lambda_1\lambda_2)^3}+\frac{2\omega_b(\lambda_1+\lambda_2)}{(\lambda_1\lambda_2)^2} \nonumber \\
&&+2\Pi_{+-}\frac{(\omega_b+\omega_f)(\lambda_1+\lambda_2)}{\Pi_{-+}(\lambda_1\lambda_2)^2}
\end{eqnarray}
The conditional randomness paremeters (eq.~(\ref{eq-for30a})) are plotted 
in the figs.~\ref{fig-2new} and \ref{fig-3new} against ATP concentration for 
different values of $\omega_b$ and $\omega_f$ respectively. The non monotonic 
variation of conditional randomness parameters with ATP concentration changes 
to a monotonic decrease (see fig.~\ref{fig-3new}) when the magnitude of 
$\omega_b$ is sufficiently high. Variation of the randomness parameter 
indicates the changes in the number of rate-limiting steps. Any value of 
the randomness parameter higher than unity may, at first sight, appear 
counter-intuitive. But, this is quite common for systems with branched 
mechano-chemical kinetics.

\section{Summary and conclusion}
\label{sec-sum-concl-kif1asingle}

Theoretical calculation of the dwell time distribution of two-headed 
conventional kinesin motors has been reported earlier \cite{valleriani}. 
In this paper we have derived an exact analytical expression for the 
distribution of dwell times of single-headed kinesin motors KIF1A at 
each binding site during a processive run on a microtubule. We have 
used the NOSC model \cite{nosc} for single-headed KIF1A motors to 
derive our results. Since both forward and backward steps of this 
motor are possible, we have also defined {\it conditional} dwell 
times and calculated their distributions analytically. The experimentally 
measured dwell time distributions for some of the other processive 
motors, like conventional kinesin \cite{asbury}, myosin-V \cite{pierobon}, 
dynein \cite{reck}, ribosome \cite{wen}, etc., have been reported in 
the literature. To the best of our knowledge, the dwell time distribution 
of KIF1A has not been reported so far; we hope our theoretical prediction 
will stimulate experimental investigations.

\acknowledgements
This work is suppoted by a research grant from CSIR (India). 
AG thanks UGC (India) for a senior research fellowship.

\end{document}